# Variational Approach for Fractional Partial Differential Equations

**Guo-cheng Wu**
Modern Textile Institute, Donghua University, 1882 Yan-an Xilu Road,
Shanghai 200051, PR China
Email: wuguocheng2002@yahoo.com.cn

**Abstract**

Fractional variational approach has gained much attention in recent years. There are famous fractional derivatives such as Caputo derivative, Riesz derivative and Riemann-Liouville derivative. Several versions of fractional variational principles are proposed. However, it becomes difficult to apply the existing fractional variational theories to fractional differential models, due to the definitions of fractional variational derivatives which not only contain the left fractional derivatives but also appear right ones. In this paper, a new definition of fractional variational derivative is introduced by using a modified Riemann-Liouville derivative and the fractional Euler-Lagrange principle is established for fractional partial differential equations.

## 1. Introduction

Any physical laws in continuous mechanics can be expressed as variational principles, which lead to differential equations. Due to dramatically development of nanotechnology and quantum mechanics, fractional calculus has been caught much attention for it can exactly describe many phenomena in discontinuous space-time. For example, fractional calculus can be applied to the Brown motion [1], an anomalous diffusion [2], transportation in porous media [3] et al. The question is now arising. Can a system of fractional equations admit a variational principle? Can a physical law in fractal space-time be derived from an action-like principle?

Recently, O.P. Argwal and D. Baleanu et al. [4- 7] defined a fractional variational principle,

$$J[y]=\int_a^b F(x, y, {}_a^C D_x^\alpha y)dx,\ \ 0<\alpha\le 1. \tag{1}$$

where $a \le x \le b,\quad y(a)=y_a,\quad y(b)=y_b$ and ${}_a^C D_x^\alpha$ is left Caputo derivative of $\alpha$ order. The corresponding variational formulation was given by

$$\frac{\partial F}{\partial y} - {}_x^C D_b^\alpha \left(\frac{\partial F}{\partial {}_a^C D_x^\alpha y}\right)=0. \tag{2}$$

However, the problem has not yet completely solved.

For example, consider the fractional equation with Caputo derivative

$$D_x^{(2\alpha)} u + u = 0, \tag{3}$$

where $D^{(2\alpha)}$ is just a notation defined by ${}_a D_x^\alpha \ {}_a D_x^\alpha$. We aim at search for a fractional Lagrangian, $L$, so that Eq. (2) can be equivalently written in the form [4]

$$\frac{\partial L}{\partial y} - {}_x D_b^\alpha \left(\frac{\partial L}{\partial {}_a D_x^\alpha \eta}\right)=0. \tag{4}$$

Obviously, we cannot find the fractional Lagrangian here.

Most recently, a new modified Riemann-Liouville left derivative is proposed by G. Jumarie [8]. The definition of the modified Riemann-Liouville derivative is not required to satisfy higher integer-order derivative than $\alpha$ . Secondly, $\alpha^{\text{th}}$ derivative of a constant is zero. For these merits, the modified derivative was successfully applied in the probability calculus [9], fractional Laplace problems [10], fractional variational calculus for single-variable system [11, 12], fractional variational iteration for fractional differential equations [13] and fractional Adomian decomposition method [14].

In this study, we used the modified Riemann-Liouville derivative, present a new definition of fractional variational derivative and establish fractional Euler-Lagrange equations for time-space fractional Burgers equation and KdV equation.

**2. Modified Riemann-Liouville Derivative**

In this section, we will introduce the definition of the modified Riemann Liouville derivative. Assume $f: R \to R, x \to f(x)$ denote a continuous (but not necessarily differentiable) function and let the partition $h > 0$ of an interval [0, 1]. Through the fractional Riemann Liouville integral

$$ {}_0I_x^{\alpha} f(x) = \frac{1}{\Gamma(\alpha)} \int_0^x (x-\xi)^{\alpha-1} f(\xi) d\xi, \alpha > 0. \tag{5} $$

With the definition of fractional variational derivative for fractional differential equations [11, 12], there is an equivalent form of the fractional integrals w.r.t $(d\xi)^{\alpha}$ [10],

$$ {}_0I_x^{\alpha} f(x) = \frac{1}{\Gamma(\alpha+1)} \int_0^x f(\xi)(d\xi)^{\alpha}, 0 < \alpha < 1. $$

The modified Riemann-Liouville derivative is defined as

$$ {}_0D_x^{\alpha} f(x) = \frac{1}{\Gamma(1-\alpha)} \frac{d}{dx} \int_0^x (x-\xi)^{-\alpha} (f(\xi) - f(0)) d\xi, \tag{6} $$

where $x \in [0,1]$, and $0 < \alpha < 1$.

The detail properties of the fractional calculus can be found in Ref. [15].

**3. Fractional Variational Principle For Fractional Differential Equations**

G. Jumarie proposed the fractional variational derivative for the single-variable system [11]

$$ \frac{\partial L}{\partial y} - {}_aD_x^{\alpha} \left( \frac{\partial L}{\partial\, {}_aD_x^{\alpha} y} \right) = 0, $$

where $L$ is the fractional Lagrange functionl. With this definition, the following two examples are illustrated the fractional variational approach for fractional differential equations.

Example 1. The classical Lagrangian is

$$ L = \frac{1}{2} \left( \frac{d\theta}{dt} \right)^2 - \frac{1}{2} mgl\theta, \tag{7} $$

where $\theta$ denotes the angular coordinate, $m$ is the mass, $g$ is the acceleration of gravity and $l$ is the pendulum length.

The possible fractional Lagrangian for this system has the form [6]

$$ L = \frac{1}{2} ({}_aD_t^{\alpha}\theta)^2 - \frac{1}{2} mgl\theta^2. \tag{8} $$

Then the fractional Euler-Lagrange equation can be calculated as

$$ {}_aD_t^{2\alpha}\theta - mgl\theta = 0. \tag{9} $$

If $\alpha \to 1$, we have

$$ \theta'' - mgl\theta = 0, \tag{10} $$

which is a simple oscillator equation.

Example 2. We consider the inverse problem of one possible fractionalized mathematical pendulum

$$y^{(2\alpha)} + \sin(y) = 0, \tag{11}$$

where $D^{(2\alpha)}$ is defined by ${}_aD_t^{\alpha}\ {}_aD_t^{\alpha}$. Then the corresponding fractional Euler-Lagrange equation is calculated as

$$L = \frac{({}_aD_t^{\alpha} y)^2}{2} + \cos(y). \tag{12}$$

**3. Fractional Variational Principle in Fractal Time and Space**

Very recently, Almeida et al [16] firstly proposed a fractional calculus for multiple integrals with Jumarie's derivative. However, the fractional order of time and space may not be equivalent in fractal time-space. Similarly, we can investigate the fractional functional with respect to $(dx)^{\beta}$ and $(dt)^{\alpha}$ :

$$J[y] = \frac{1}{\Gamma(1+\alpha)\Gamma(1+\beta)} \iint_D L(t, x, y, {}_aD_t^{\alpha} y, {}_aD_x^{\beta} y)(dt)^{\alpha}(dx)^{\beta}, \tag{13}$$

where $D = (x,t)$ is a closed region: $c \le x \le d$ and $a \le t \le b$. Here $\alpha$ is the fractional order which can be explained as fractal dimension in discontinuous time or space, see Ref. [17 - 18].

Let $\Omega$ be a positively oriented, piecewise smooth, simple closed curve in the plane $R^2$, and let $D$ be the region bounded by $\Omega$. Then we can have the generalized Green Theorem,

$$\begin{aligned} &\frac{1}{\Gamma(1+\alpha)\Gamma(1+\beta)} \iint_D [Q_t^{(\alpha)}(t,x) - P_x^{(\beta)}(t,x)](dt)^{\alpha}(dx)^{\beta} \\ &= \frac{1}{\Gamma(1+\beta)} \square \quad \text{[illegible]} \quad \frac{1}{\Gamma(1+\alpha)} \square \end{aligned} \tag{14}$$

where $Q(x,y)$ and $P(x,y)$ are continuous (but not necessarily differentiable) functions defined on the region $D$.

Assume $y^*$ is the extremal surface of the functional $J$ and $y$ is the near-surface,

$$y = y^* + \varepsilon\eta(x,t) \tag{15}$$

where $\eta(x,t)$ takes given values on the boundary $\Omega$ of $D$.

Set

$$\eta(x,t)|_{\Omega} = 0, \tag{16}$$

for simplicity, then

$$\delta y = \varepsilon\eta(x,t)|_{\Omega} = 0. \tag{17}$$

${}_aD_t^{\alpha}$ and ${}_cD_x^{\beta}$ are line operators, so we can have the following relations

$$\begin{cases} \delta({}_cD_x^{\beta} y) = {}_cD_x^{\beta} \delta y, \\ \delta({}_aD_t^{\alpha} y) = {}_aD_t^{\alpha} \delta y. \end{cases} \tag{18}$$

Making Eq. (13) stationary with respect to $y$, we have

$$\begin{aligned} \delta J &= \frac{1}{\Gamma(1+\alpha)\Gamma(1+\beta)} \iint_D [\frac{\partial L}{\partial y}\delta y + \frac{\partial L}{\partial_a D_t^{\alpha} y}\delta y^{(\alpha)} + \frac{\partial L}{\partial_c D_x^{\beta} y}\delta y^{(\beta)}](dt)^{\alpha}(dx)^{\beta} \\ &= \frac{1}{\Gamma(1+\alpha)\Gamma(1+\beta)} \iint_D [\frac{\partial L}{\partial y}\delta y + \frac{\partial L}{\partial_a D_t^{\alpha} y}(\delta y)^{(\alpha)} + \frac{\partial L}{\partial_c D_x^{\beta} y}(\delta y)^{(\beta)}](dt)^{\alpha}(dx)^{\beta}. \end{aligned} \tag{19}$$

With the fractional Leibniz law [9 - 11], we can derive

$$\begin{aligned} &\frac{\partial L}{\partial_a D_t^{\alpha} y}\delta y^{(\alpha)} + \frac{\partial L}{\partial_c D_x^{\beta} y}\delta y^{(\beta)} \\ &= [{}_aD_t^{\alpha}(\frac{\partial L}{\partial_a D_t^{\alpha} y}\delta y) + {}_cD_x^{\beta}(\frac{\partial L}{\partial_c D_x^{\beta} y}\delta y)] - \delta y[{}_aD_t^{\alpha}\frac{\partial L}{\partial_a D_t^{\alpha} y} + {}_cD_x^{\beta}\frac{\partial L}{\partial_c D_x^{\beta} y}]. \end{aligned} \tag{20}$$

Substitute Eq. (20) into Eq. (19) results in

$$\begin{aligned} \delta J &= \frac{1}{\Gamma(1+\alpha)\Gamma(1+\beta)} \iint_a^b \delta y[\frac{\partial L}{\partial y} - {}_aD_t^{\alpha}\frac{\partial L}{\partial_a D_t^{\alpha} y} - {}_cD_x^{\beta}\frac{\partial L}{\partial_c D_x^{\beta} y}](dt)^{\alpha}(dx)^{\beta} \\ &+ \frac{1}{\Gamma(1+\alpha)\Gamma(1+\beta)} \iint_a^b [{}_aD_t^{\alpha}(\frac{\partial L}{\partial_a D_t^{\alpha} y}\delta y) + {}_cD_x^{\beta}(\frac{\partial L}{\partial_c D_x^{\beta} y}\delta y)](dt)^{\alpha}(dx)^{\beta} \\ &= \delta J_1 + \delta J_2. \end{aligned}$$

Using the fractional Green theorem, Eq. (14), we can note

$$\begin{aligned} \delta J_2 &= \frac{1}{\Gamma(1+\alpha)\Gamma(1+\beta)} \iint_D [{}_aD_t^{\alpha}(\frac{\partial L}{\partial_a D_t^{\alpha} y}\delta y) + {}_cD_x^{\beta}(\frac{\partial L}{\partial_c D_x^{\beta} y}\delta y)](dt)^{\alpha}(dx)^{\beta} \\ &= \frac{1}{\Gamma(1+\beta)} \square \; \frac{\partial L}{\partial_a D_t^{\alpha} y} \cdots dx)^{\beta} - \frac{1}{\Gamma(1+\alpha)}\frac{\partial L}{\partial_c D_x^{\beta} y}\delta y(dt)^{\alpha}] \\ &= 0. \end{aligned} \tag{21}$$

As a result, we can obtain

$$\delta J = \frac{1}{\Gamma(1+\alpha)\Gamma(1+\beta)} \iint_D \delta y[\frac{\partial L}{\partial y} - {}_aD_t^{\alpha}\frac{\partial L}{\partial_a D_t^{\alpha} y} - {}_cD_x^{\beta}\frac{\partial L}{\partial_c D_x^{\beta} y}](dt)^{\alpha}(dx)^{\beta}.$$

Since $\delta y$ is arbitrary in the domain $D$, the condition $\delta J = 0$ implies that

$$\frac{\partial L}{\partial y} - {}_aD_t^{\alpha}\frac{\partial L}{\partial_a D_t^{\alpha} y} - {}_cD_x^{\beta}\frac{\partial L}{\partial_c D_x^{\beta} y} = 0. \tag{22}$$

Generally, when the functional $J$ contains higher order fractional derivative terms, its Euler-Lagrange equation now can be written in the form

$$\frac{\partial L}{\partial u}+\sum_{m}(-1)^{m}{}_{a}D_{t}^{m\alpha}\left(\frac{\partial L}{\partial_{a}D_{t}^{m\alpha}u}\right)+\sum_{n}(-1)^{n}{}_{c}D_{x}^{n\beta}\left(\frac{\partial L}{\partial_{c}D_{x}^{n\beta}u}\right)=0,\ 0<\alpha,\ \beta<1, \tag{23}$$

where $m$ and $n$ are positive integers.

If $\alpha=\beta=1$, Eq. (23) can turn out to be the Euler-Lagrange equation in usual sense.

## 4. Variational Approach for Fractional Partial Differential Equations

The semi-inverse method or called He's variational approach was first proposed in 1997 to search for variational formulations directly from governing equations and boundary/initial conditions [19]. The method has been used by many researchers to establish various variational principles in different fields, for examples, Adali first established a variational principle for a multi-walled carbon nanotube [20]; Zheng and Wang found a generalized variational principle in special relativity [21]. He and Lee proposed variation formulas for differential-difference equations [22]. In this section, we will further extend this method to fractional partial differential equations. In order to illustrate our fractional semi-inverse method, we now establish the fractional Euler Lagrange equations for the time-space fractional Bugers equation and KdV equation

$$u_{t}^{(\alpha)}=uu_{x}^{(\beta)}+f(x,t),\ 0<t,x,\ 0<\alpha,\beta\le 1. \tag{24}$$

We firstly introduce a function, $\varphi$, defined as

$$\begin{cases} {}_{0}D_{x}^{\beta}\varphi=u, \\ {}_{0}D_{t}^{\alpha}\varphi=\dfrac{u^{2}}{2}+F(x,t), \end{cases} \tag{25}$$

where ${}_{0}D_{x}^{\beta}F(x,t)=f(x,t)$.

Construct the following functional

$$L=u_{0}D_{t}^{\alpha}\varphi-\left(\frac{u^{2}}{2}+F(x,t)\right){}_{0}D_{x}^{\beta}\varphi+G. \tag{26}$$

Then the Euler-Lagrange equation with respect to $u$ is

$${}_{0}D_{t}^{\alpha}\varphi-u_{0}D_{x}^{\beta}\varphi+\frac{\delta G}{\delta u}=0, \tag{27}$$

where $\delta G/\delta u$ is called the fractional variational derivative defined as Eq. (22) or (23)

We can readily find

$$G=\frac{u^{3}}{6}-F(x,t)u. \tag{28}$$

Constraining the fractional variational principle, Eq. (26), by Eq. (25), we can obtain a constrained variational principle

$$J[\varphi]=\frac{1}{\Gamma(1+\alpha)\Gamma(1+\beta)}\iint_D\left(\frac{u^3}{6}-F(x,t)u\right)(dt)^{\alpha}(dx)^{\beta}. \tag{29}$$

If $\beta=1$, Eq. (29) can turn out to be a constrained variational principle for the Example. 1. [23].

The possible fractional KdV equation can be given as

$$_0D_t^{\alpha}u=6u_0D_x^{\beta}u+{}_0D_x^{3\beta}u={}_0D_x^{\beta}(3u^2+{}_0D_x^{2\beta}u), \tag{30}$$

where $_0D_x^{2\beta}$ is just a notation and defined as $_0D_x^{2\beta}={}_0D_x^{\beta}{}_0D_x^{\beta}$ and $0<\alpha,\ \beta\le 1$.

Introduce a function, $\varphi$, defined as

$$\begin{cases} _0D_x^{\beta}\varphi=u, \\ _0D_t^{\alpha}\varphi=3u^2+{}_0D_x^{2\beta}u, \end{cases} \tag{31}$$

so that Eq. (30) is automatically satisfied.

Similarly, a trial-Lagrangian can be constructed in the form

$$L=u_0D_t^{\alpha}\varphi-(3u^2+{}_0D_x^{2\beta}u)\ {}_0D_x^{\beta}\varphi+F, \tag{32}$$

where $F$ is an unknown function of $u$ and/or its derivatives. The advantage of the trial-Lagrangian, Eq. (26), lies on the fact that the fractional Euler-Lagrange equation with respect to $\varphi$ is Eq. (30).

The generalized Euler-Lagrange equation with respect to $u$ is

$$_0D_t^{\alpha}\varphi-6u_0D_x^{\beta}\varphi-{}_0D_x^{3\beta}\varphi+\frac{\delta F}{\delta u}=0, \tag{33}$$

Therefore, we can obtain

$$\frac{\delta F}{\delta u}=-{}_0D_t^{\alpha}\varphi+6u_0D_x^{\beta}\varphi+{}_0D_x^{3\beta}\varphi=-3u^2-{}_0D_x^{2\beta}u+6u^2+{}_0D_x^{2\beta}u=3u^2,$$

from which we can identify $F$ in the form

$$F=u^3. \tag{34}$$

We, therefore, obtain the following fractional variational principle for Eq. (30)

$$J[u,\varphi]=\frac{1}{\Gamma(1+\alpha)\Gamma(1+\beta)}\iint_D\left\{u_0D_t^{\alpha}\varphi-(3u^2+{}_0D_x^{2\beta}u)\ {}_0D_x^{\beta}\varphi+u^3\right\}(dt)^{\alpha}(dx)^{\beta}. \tag{35}$$

## 5. Conclusion

Fractional variational calculus has gained considerable attention during the last decade due to their various applications in several areas of science and engineering. However, in the problem of the variational calculus, it is difficult to obtain a general Lagrange function using classical approaches. The fractional variational principle given in this study provides a universal tool to investigating fractional partial differential equations and other fractional nonlinear techniques depended on the variational formulas now can be consider to further extend. We will discuss such work in future.